\documentclass{aa}  

\usepackage{graphicx}
\usepackage{txfonts}
\usepackage{booktabs}
\usepackage[outercaption]{sidecap}
\usepackage[colorlinks=true,
            linkcolor=blue,
            urlcolor=blue,
            citecolor=blue]{hyperref}

\newcommand{\Teff}{\ensuremath{T_\mathrm{eff}}}
\newcommand{\logg}{\ensuremath{\log g}}
\newcommand{\vrad}{\ensuremath{v_\mathrm{rad}}}
\newcommand{\vsini}{\ensuremath{v_\mathrm{rot}\sin i}}
\newcommand{\lognHE}{\ensuremath{\log (n_\mathrm{He}/n_\mathrm{H})}}

\newcommand{\teff}{$T_\mathrm{eff}$}

\begin{document}

   \title{Stars on the ascending helium giant branch  }

   \subtitle{I. From white dwarf merger to helium giant: the evolutionary state of the rapidly rotating hot subdwarf HE~1518-0948}

   \author{M.~Pritzkuleit
          \inst{1}
          \and
          M.~Dorsch
          \inst{1}
          \and 
          M. M.~Miller Bertolami
          \inst{2,3}
          \and
          S.~Geier
          \inst{1}
          \and
          C.~W.~Bradshaw
          \inst{4}
          \and
          H.~Dawson
          \inst{1}
          }

   \institute{Institut für Physik und Astronomie, Universität Potsdam,
              Karl-Liebknecht-Str. 24/25, D-14476 Potsdam, Germany\\
              \email{maxpri@astro.physik.uni-potsdam.de}
      \and
            Instituto de Astrofísica de La Plata, Consejo Nacional de Investigaciones Científicas y Técnicas Avenida Centenario (Paseo del Bosque) S/N, B1900FWA La Plata Argentina
      \and  
            Facultad de Ciencias Astronómicas y Geofísicas, Universidad Nacional de La Plata Avenida Centenario (Paseo del Bosque) S/N, B1900FWA La Plata Argentina  
      \and 
             Hamburger Sternwarte, University of Hamburg, Gojenbergsweg 112, D-21029 Hamburg, Germany
             }
   \date{}

 
  \abstract
   {Hot subdwarf stars with masses above $0.8~M_\odot$ ascend the helium giant branch after the end of core helium burning, before entering the white dwarf cooling track or exploding as type Ib/c supernovae. Such massive helium stars are typically expected to form through the stripping of an intermediate mass star by a binary companion. Even after being stripped, the stars are expected to retain a detectable amount of hydrogen but there is also the class of extreme helium-rich hot subdwarfs (He-sdOs), which show no, or only very weak, traces of hydrogen in their spectra. Several evolutionary channels have been proposed to explain their formation, but their extremely low binary fraction suggests that they are either created through single-star evolution triggered by a late hot flash in a low-mass red giant or the merger of two helium white dwarfs (WDs). Most He-sdOs are located close to the helium zero-age main sequence, while a small number exhibit much lower surface gravities, indicating inflated radii. Whether these objects are evolutionarily connected to the main He-sdO population remains unclear.

   In this work, we analyse the extreme helium-rich, low-surface-gravity sdO HE~1518–0948 (HE~1518) through a detailed spectroscopic study of optical and ultraviolet data. We measure an effective temperature of $52~000\pm2500$~K, a surface gravity {\logg} of $4.64\pm0.15$~dex, 
   an upper limit for the hydrogen abundance of \lognHE$~<2.5$, and an exceptionally high projected rotational velocity of $90\pm20~\mathrm{km~s^{-1}}$, significantly larger than that of most known He-sdOs. The star is found to belong kinematically to the Galactic halo, consistent with the very low metallicity derived from our abundance analysis. A comparison with evolutionary models indicates that HE~1518 is the product of a massive double helium white dwarf merger and is currently undergoing helium shell burning while ascending the helium giant branch. This makes HE~1518 one of only a few known objects located in this sparsely populated region of the Hertzsprung–Russell diagram. Such stars provide valuable laboratories for studying the evolution of massive hot subdwarfs beyond core helium burning, and their high luminosities allow them to be detected at large distances. }


   \maketitle
%
\section{Introduction}
Hot subdwarf O (sdO) and B (sdB) stars are evolved core or shell helium-burning stars that lost most of their hydrogen-rich envelope \citep[for reviews, see][]{Heber2016, Heber2026}. The subclass of helium-rich sdOs (He-sdO) shows only very weak or even no hydrogen traces in their spectra. Radial velocity (RV) surveys of hydrogen-rich sdBs reported binary fractions between 30-70\% \citep{Maxted2001, Copperwheat2011, Kawka2015, Geier2022, He2025}, and \citet{Pelisoli2020} provide observational evidence that binary interactions are likely always required for their formation. The main proposed formation channels are a stable Roche lobe overflow or a common envelope phase of low-mass stars ($\lesssim1.8~M_\odot$) at the tip of the red giant branch evolution. In case of intermediate-mass stars ($1.8~M_\odot<M\lesssim10~M_\odot$), it might already happen after the end of core hydrogen burning \citep{Han2002, Han2003, Goetberg2018}. 

In contrast, observational studies indicate that helium-rich sdOs are rarely found in binaries. \citet{Snowdon2025} performed a binarity search among the extreme and intermediate helium-rich hot subdwarfs observed in the SALT survey of hydrogen-deficient stars \citep{Jeffery2023, Jeffery2026}. The authors found that all extreme helium-rich objects in their sample were single, which is in clear contrast to the hydrogen-rich objects \citep[see also][]{Geier2022, He2025}. Thus, there is strong evidence that the helium-rich hot subdwarfs originate from evolutionary channels different than the envelope stripping scenarios proposed for hydrogen-rich sdBs.  

Most He-sdOs have effective temperatures ($\mathbf{T_\mathrm{eff}}$) between 40 and 55~kK and surface gravity ({\logg}) values between 5.5 and 6.0~dex \citep[e.g.][]{Stroeer2007, Nemeth2012, Luo2021, Latour2026}. They are divided into intermediate helium-rich objects (iHe-sdOBs), characterised by atmospheric helium abundances in the range $-1.2~<$~\lognHE~$<~0.6$,  and extreme helium-rich hot subdwarfs (eHe-sdOs), which have \lognHE~$>~0.6$ \citep{Latour2026, Dawson2026, Heber2026b}. The eHe-sdOs can further be divided into three subclasses depending on the amount of nitrogen and carbon visible in their spectra, as either C-rich, C\&N-rich, or N-rich but C\&O-poor \citep{Stroeer2007, Hirsch2009}. 

These abundance patterns are typically explained by the merger of two helium white dwarfs \citep[WD;][]{Webbink1984, Zhang2012}. The C-rich stars are created through a (hot) fast merger where some of the helium ignites and burns into carbon, which enriches the atmosphere (C-type). In this process, nitrogen is destroyed through $\alpha$-capture and transformed into oxygen and neon. However, carbon enhancement could also result from a late hot flash \citep{Bertolami2008, Battich2018}. In this scenario \citep{Lanz2004}, a red giant experiences the helium flash after losing most of its envelope. Recent simulations from \citet{Battich2025} show that this event can produce the conditions required for $i$-process nucleosynthesis. This mechanism could explain the extreme overabundance of heavy metals observed in several intermediate helium-rich hot subdwarfs \citep{Dorsch2020, Dorsch2021}. 

In case of a slow He-WD+He-WD merger, no burning occurs, leaving surface material enriched in nitrogen as created by the CNO burning in the progenitor stars, observed as an N-type He-sdO. Another proposed channel to  form He-sdOs of the N-type is the stripping of a H-rich sdO/B by a close compact companion \citep{Rajamuthukumar2025}. Progenitor binaries, some of them with accretion disks, have been found \citep[e.g.][]{Kupfer2020b, Kupfer2020a, Stringer2023} as well as the N-type He-sdO US~708, which is thought to be the stripped donor ejected by a supernova type Ia \citep{Geier2015}. However, the low binary fraction of eHe-sdOs implies that this cannot be the dominant formation channel for this subtype. 

Helium-rich hot subdwarfs enriched in carbon and nitrogen (CN-type) are predicted to form when part of the material is accreted rapidly during a WD merger, while the remainder is accreted more slowly from a surrounding disk, thus increasing the nitrogen abundance \citep{Zhang2012}. Also the late hot flasher scenario might result in CN-type He-sdOs \citep{Bertolami2008, Battich2018}. Stars with a similar surface composition, but very different internal structure, can also be created during a merger of a carbon-oxygen (C/O) WD and a He-WD \citep{Jeffery2011}.
The remnant initially evolves into a helium giant, observed as an R Coronae Borealis (R~CrB) star, and subsequently appears as a very luminous, helium giant classified as an extreme helium  (EHe) star \citep{Jeffery2008} and later O(He) star \citep{Reindl2014}, before descending onto the white dwarf cooling track.

A few He-sdOs have surface gravities below 5.0~dex and are therefore located far above the helium zero age main sequence (He-ZAMS). These objects may follow an evolutionary pathway different from that of the bulk of the He-sdO population, potentially representing a distinct subgroup. 

Here, we present a detailed analysis of the low-gravity hot subdwarf HE~1518$-$0948 (hereafter HE~1518) that was first classified by \citet{Moehler1990} as He-sdO. \citet{Stroeer2007} reported an unusually low surface gravity but were unable to model the spectrum sufficiently because it was outside their grid boundaries. In this paper, we present a detailed analysis of the star by combining available spectroscopic, photometric, and astrometric data to derive its atmospheric parameters and elemental abundances (Sects.~\ref{specan}–\ref{kin_sect}). Based on these results and a comparison with evolutionary tracks, we discuss its evolutionary history and future, as well as its enrichment with nuclear-processed material compared to previously analysed He-sdOs (Sect.~\ref{discussion}).

\section{Observations}
Spectroscopic observations of HE~1518 span more than 20 years and cover wavelengths from the ultraviolet (UV) to the infrared. The spectroscopic analysis presented here is based on two high-resolution optical spectra obtained with the Ultraviolet and Visual Echelle Spectrograph (UVES) mounted on the Very Large Telescope (VLT), as well as UV spectra acquired with the Space Telescope Imaging Spectrograph (STIS) on board the Hubble Space Telescope (HST).

The UVES spectra were obtained in the course of the ESO Supernova Ia Progenitor SurveY (SPY; \citealt{Napiwotzki2020}). They have a resolving power of $R \simeq 20~000$ and cover a wide wavelength range from $3300$ to $6650~\AA$, with gaps at $4500$–$4600~\AA$ and $5600$–$5700~\AA$. For the spectroscopic analysis both spectra were co-added resulting in an average signal-to-noise ratio (S/N) of 36.

The STIS spectra have a resolving power of $R = 45~800$, making them the highest-resolution data available for this star. They cover the wavelength range $1144$–$1729~\AA$. HE~1518 was observed during three consecutive HST orbits, and the resulting spectra were co-added to improve the S/N. Owing to the high projected rotational velocity of the star (see Section~\ref{specan}), the spectrum was further rebinned to a resolving power of ${R~=~20~000}$ without significant loss of information, yielding an average S/N of about 72.

The remaining 30 spectra were taken mostly in the course of the Massive Unseen Companions to Hot Faint Underluminous Stars from SDSS (MUCHFUSS) program \citep{Geier2011, Kupfer2015} using the Intermediate Dispersion Spectrograph (IDS) at the Isaac Newton Telescope (INT), the Intermediate-dispersion Spectrograph and Imaging System (ISIS) at the William Herschel Telescope (WHT) and the Goodman spectrograph at the Southern Astrophysical Research Telescope (SOAR). The spectra obtained with these instruments were used only to measure their radial velocities because of their lower resolving power compared to the UVES and STIS data. An overview of all spectra is given in Table~\ref{observations}.

\section{Spectral analysis and spectroscopic parameters}\label{specan}
To determine \Teff, \logg, projected rotational velocity (\vsini), radial velocity (\vrad), and elemental abundances, we fitted the UVES and STIS spectra. They were chosen for their high resolution and extensive coverage, spanning the UV to infrared range. 

As an initial estimate of the atmospheric parameters, we fitted the UVES spectrum using a grid of non-local thermodynamic equilibrium (NLTE) model atmospheres computed with TLUSTY/SYNSPEC \citep{Hubeny2017a}. The grid spans effective temperatures of 26250 to 57500~K, surface gravities of $4.25 \le \log g \le 6.5$, and helium abundances of $-1.75~\le$~\lognHE~$\le~4.00$. The models include full line blanketing\footnote{The grid assumes solar abundances, except for C (2$\times$), O (0.1$\times$), Ne (2$\times$), Fe (1.5$\times$), and Ni (10$\times$ solar); enhanced Fe and Ni compensate for the omission of other iron-group elements.} and assume plane-parallel, homogeneous, hydrostatic atmospheres; further details are given in \citet{Dorsch2024}. Spectral fitting was performed with the Interactive Spectral Interpretation System (ISIS; \citealt{Houck2000}), which applies a $\chi^2$ minimisation to hydrogen and helium line profiles following the method of \citet{Irrgang2014} to determine the best-fitting parameters.

An initial modelling attempt by \citet{Stroeer2007} placed the star outside the boundaries of their grid. They suggest an effective temperature of around 60~000~K and a surface gravity {\logg} below 4.8~dex. Our analysis yields a temperature of $52~000\pm2500$~K, a surface gravity of $4.64\pm0.15$~dex and a projected rotational velocity {\vsini} of $90\pm20~\mathrm{km~s}^{-1}$. 

\begin{figure*}
\centering
\includegraphics[width=0.99\textwidth]{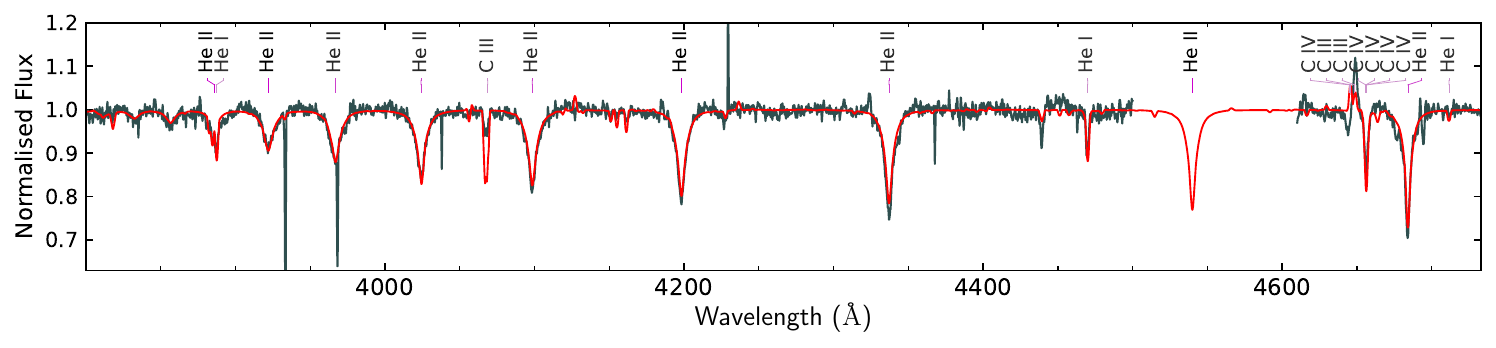}\\[-19pt]
\includegraphics[width=0.99\textwidth]{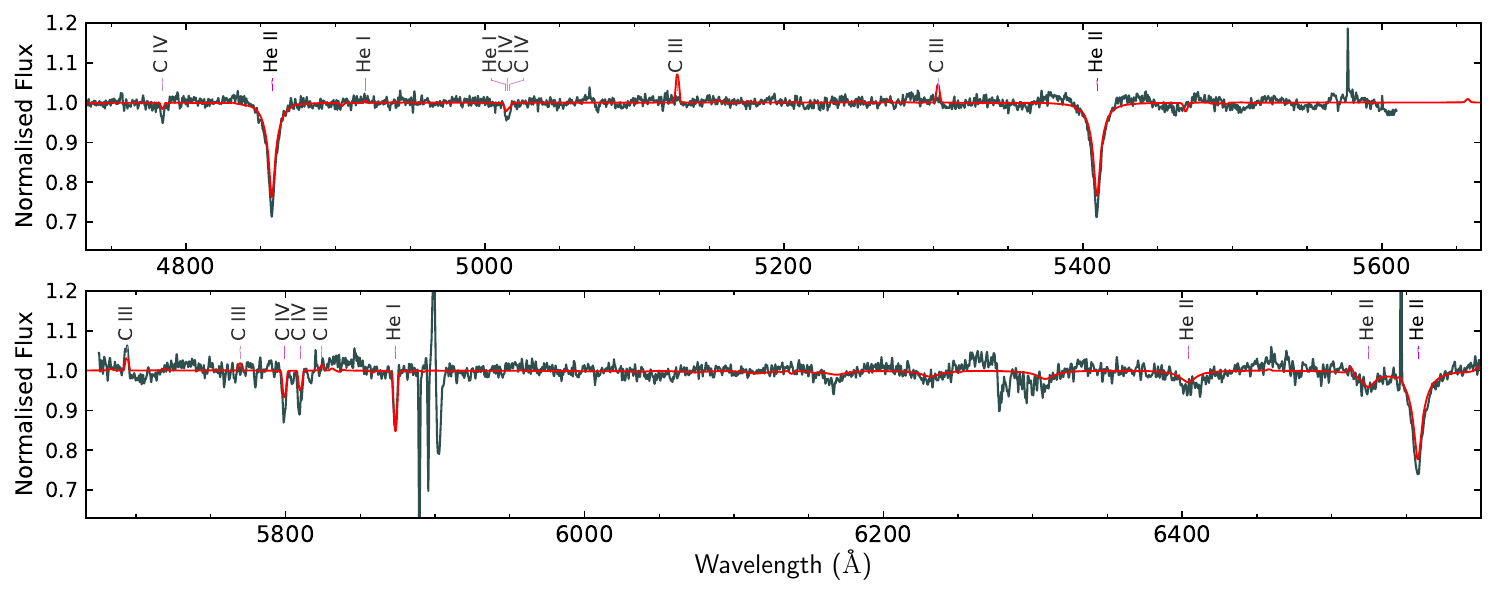}
\caption{Spectral fit of the optical UVES spectrum shown in black and overplotted is the model in red.}
\label{SpecfitOPT}
\end{figure*}

Despite including line blanketing and non-LTE effects, the models underpredict the strengths of the \ion{He}{ii} lines as can be seen in Fig.~\ref{SpecfitOPT}. This persistent discrepancy, known as the Balmer/He~\textsc{ii} line problem, reflects remaining limitations in the modelling of hot stellar atmospheres and has been noted in analyses of hot subdwarfs \citep{Lanz1997, Latour2015} and white dwarfs \citep{Werner2018}, particularly at high \teff. Therefore, the effective temperature is constrained primarily by the \ion{He}{i} lines and the UV C~\textsc{iii}/\textsc{iv} features (see Sect.~\ref{Sect_abu}). In our models, this issue becomes apparent at temperatures of about 46~000~K. To compensate for this effect, $\chi^2$ fitting tends to favour models with higher hydrogen abundances, as the overlap between the \ion{He}{ii} Pickering and H Balmer series partially fills in the affected lines. To avoid this bias, we set \lognHE$~=3.5\pm0.5$ and enforce the H-poor solution, although small quantities of hydrogen may lie below \lognHE~$<2.5$ which we determine as our detection threshold. Above this limit, the hydrogen emission in the H$\alpha$ line becomes visible.

\section{Abundances}
\label{Sect_abu}
The values determined in Sect.~\ref{specan} served as initial parameters for constructing a refined TLUSTY/SYNSPEC model, which was then applied to the STIS and UVES spectra. The parameters \teff, \logg, and {\lognHE} were kept fixed. The elements C, N, O, Ne, Mg, Si, P, S, Fe, and Ni are treated in NLTE whereas Al, Ar, Ti, Cr, Mn, Co, Zn, Ge, Sr, Zr and Pb are in LTE. The abundances of these elements were initially determined by visual inspection of the STIS spectrum. For elements lacking visible absorption features, an upper limit for their abundance was estimated by gradually increasing the abundance until lines become detectable. Using these abundances, we created a new grid with SYNSPEC, fixing the effective temperature and surface gravity, while varying the abundances of elements heavier than helium by $\pm 0.9$ dex in steps of 0.3 dex. A second fitting procedure was performed using ISIS, and the resulting element abundances were used to calculate a final TLUSTY model. Because of the different opacities between the initial and final models, we examined the UV C~\textsc{iii}/\textsc{iv} lines to assess whether an update of \teff\ was required. No adjustment was found to be necessary.
The final model is shown in the optical range in Fig.\ \ref{SpecfitOPT}. 

The optical spectrum is dominated by He \textsc{ii} and C \textsc{iii}/\textsc{iv} features. A prominent characteristic is the C \textsc{iii} emission line near 4650 \AA, bracketed by two C \textsc{iv} absorption lines (Fig.~\ref{SpecfitOPT}). A similar configuration has been reported in the SALT survey of hydrogen-deficient stars \citep{Jeffery2021}, particularly among high-\teff, low-\logg\ He-sdO stars such as HE1518. Although our non-LTE models predict emission in the \ion{C}{iii} 4650 and 5772 \AA\ lines, the overall fit to optical metal lines remains poor. Similar difficulties with optical carbon lines have been shown previously \citep[e.g.][]{Luo2024} and appear to be associated with the large population fraction of the high-lying energy levels involved in these transitions, which are hard to model. In contrast, the low-lying UV carbon lines are not affected and are well reproduced, besides wind features in \ion{C}{iv} 1548, 1551\,\AA. 

\begin{figure}
\centering
\includegraphics[width=1.0\linewidth]{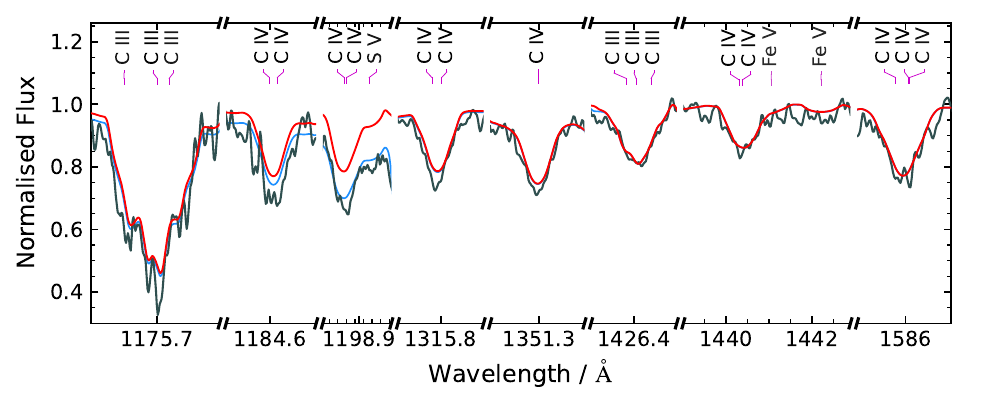}\\[-12.5pt]
\includegraphics[width=1.0\linewidth]{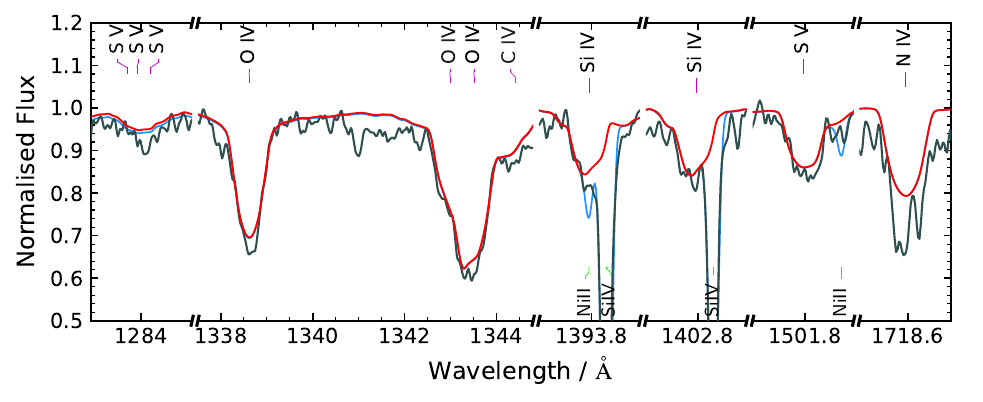}
\caption{Selected UV lines of carbon, nitrogen, oxygen, silicon, sulfur, and iron. The black line is the observation, red shows the model and light blue further includes interstellar lines.}
\label{SpecfitUV}
\end{figure}

The STIS spectrum shows additional lines of N, O, Si, S and Fe in comparison to the optical spectrum allowing us to determine the metallicity of the star. In contrast to the optical spectrum, the carbon lines in the UV are well reproduced (Fig.~\ref{SpecfitUV}), highlighting the importance of UV spectroscopy for reliable abundance determinations in such hot stars.

\begin{figure}
\centering
\includegraphics[width=1.0\linewidth]{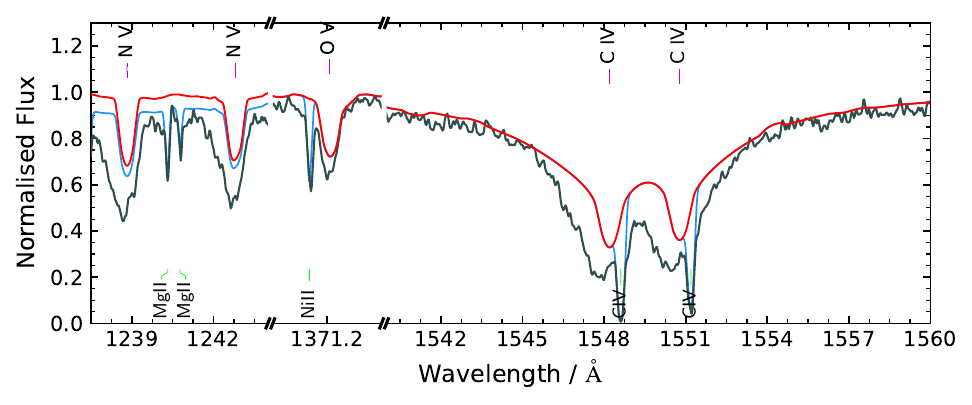}
\caption{UV lines of C, N and O affected by the stellar wind of HE~1518. The meaning of the colours is the same as in Fig.~\ref{SpecfitUV}.}
\label{SpecfitWind}
\end{figure}

The star also has a noticeable stellar wind that creates the asymmetric shape of the N~\textsc{v}, O~\textsc{v} and C~\textsc{iv} lines shown in Fig.~\ref{SpecfitWind}. From this plot, we can also see that HE~1518 is ionising its surrounding interstellar medium (ISM), creating the sharp absorption components in the C~\textsc{iv} lines. 

\begin{figure}
  \resizebox{\hsize}{!}{\includegraphics{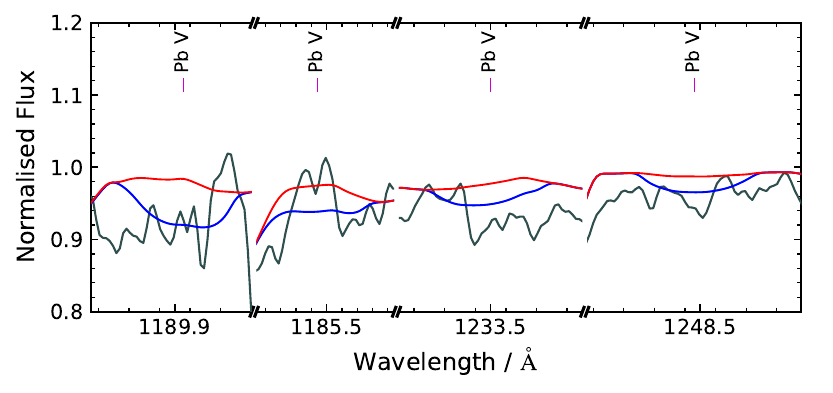}}
  \caption{Abundance comparison for lead. The blue model includes all abundances as measured from the $\chi^2$ fitting while the red line is the model without lead. Because there is no unambiguous line identification, we consider this only as upper limit. This technique was also used for all other elements for which we provide upper limits.}
  \label{Example_abu}
\end{figure}

With an iron abundance roughly 40 times lower than the solar abundance, the star is very metal poor. Because of the large rotational velocity the iron lines are shallow but remain clearly detectable. We also examined Ne, Mg, P, Al, Ar, Ti, Cr, Mn, Co, Ni, Zn, Ge, Sr, Zr, and Pb in detail by conducting a visual inspection of their strongest features, as listed in Table \ref{elements}. For each of these elements, a second model was calculated with the abundance set to zero. By comparing these two models, as exemplified for lead in Fig.~\ref{Example_abu}, we determined whether the element was detected or if only an upper limit could be established. No unambiguous detection was obtained for these elements. The abundances and upper limits are given in Table~\ref{abundances}. Solar abundances were taken from \citet{Asplund2009}.

The uncertainties of C, N, O, Si, S, and Fe were derived by iteratively increasing and decreasing the abundances until the model no longer provided an acceptable fit to the observations.

\begin{table}
\caption{Abundances of HE~1518–0948 expressed in logarithmic units. }           
\label{abundances}      
\centering                                      
\begin{tabular}{l r r r}          
\toprule\toprule   
Element & number  & mass  & solar fraction\\ 
 & $\log (n_\mathrm{X}/\Sigma_i n_i)$  & $\log (m_\mathrm{X}/\Sigma_i m_i)$  & $\log (n_\mathrm{X}/n_{\mathrm{X},\odot})$\\ \midrule
H &  $<$ -2.41 & $<$ -3.02& $<$ -2.37\\
He &  $-0.005$ & -0.02& 1.10\\
C &  $-2.01\pm0.3$ & $-1.54\pm0.3$& 1.60\\
N &  $-4.91\pm0.5$ & $-4.37\pm0.5$ & -0.70\\
O &  $-3.18\pm0.2$ & $-2.59\pm0.2$& 0.16\\
Ne & $<$ -3.99 & $<$ -3.29& $<$ 0.12\\
Mg & $<$ -3.37 &$<$ -2.59&$<$ 1.07\\
Al & $<$ -5.50 &$<$ -4.68&$<$ 0.09\\
Si & $-5.32\pm0.5$ & $-4.49\pm0.5$ & -0.80\\
P  & $<$ -5.09 & $<$ -4.21 &$<$ 1.54 \\ 
S &  $-5.30\pm0.4$ & $-4.41\pm0.4$& -0.39\\
Ar & $<$ -5.28 &$<$ -4.29&$<$ 0.36\\
Ti & $<$ -6.18& $<$ -5.12&$<$ 0.90\\
Cr & $<$ -6.03 &$<$ -4.92&$<$ 0.37\\
Mn & $<$ -6.21 &$<$ -5.08& $<$ 0.40\\
Fe &  $-6.09\pm0.5$ & $-4.96\pm0.5$& -1.56\\  
Co & $<$ -6.71 &$<$ -5.55& $<$ 0.34\\
Ni &$<$ -6.62 & $<$ -5.47& $<$ -0.81\\
Zn & $<$ -7.52 &$<$ -6.32 &$<$ -0.05\\
Ge & $<$ -6.75 & $<$ -5.50&$<$ 1.63\\
Sr & $<$ -3.81 & $<$ -2.48& $<$ 5.36 \\
Zr & $<$ -4.33 & $<$ -2.98  & $<$ 5.13 \\
Pb & $<$ -6.48 &$<$ -4.78&$<$ 3.80\\
\bottomrule                                           
\end{tabular}
\end{table}

\section{Spectral energy distribution}
\label{sect:sed}

By calculating the spectral energy distribution (SED), mass, radius and luminosity can be obtained if the star has an accurate distance determination. However, in \textit{Gaia} Data Release 3 \citep{GaiaDR3}, the parallax of the star is 0.14~mas with an error of 0.04~mas. When applying the zero-point offset following \cite{Lindegren2021} and inflating the uncertainty using the method described in \cite{ElBadry2021}, we obtained $0.17\pm0.05~\mathrm{mas}$. Due to the large error, we did not consider the parallax to be reliable. Instead, we estimated a mass of 0.9 $M_\odot$ with an uncertainty of 0.1 $M_\odot$ by using evolutionary models and theoretical considerations. See Sect.~\ref{massdet} for details. 

To model the SED, we fixed \teff, \logg, {\lognHE} and our mass estimate, while the free fitting parameters were the angular diameter $\theta$ and the colour excess $E(44-55)$. Interstellar reddening was accounted for as described in \cite{Fitzpatrick2019}. The best-fit solution was obtained by minimising $\chi^2$ using the same model grid as in the spectroscopic analysis. Further details about the fitting procedure can be found in \cite{Heber2018}. 

The results of the spectral and SED analysis, including their $1\sigma$ uncertainties, are presented in Table~\ref{params} and the SED fit is shown in Fig.~\ref{SED_fig}. There is no visible infrared excess, allowing us to rule out a luminous cool companion star such as a red giant or subgiant, while a main sequence star would be outshone by the primary. The spectroscopic distance from our analysis is consistent with the parallax distance at the $1\sigma$ level.

\begin{table}
\caption{Atmospheric and stellar parameters of HE~1518-0948. }       
\label{params}      
\centering                                      
\begin{tabular}{l r}          
\toprule
\toprule
        \multicolumn{2}{c}{Spectral analysis} \\
\midrule
    \Teff~(K) & $52~000\pm~2500$ \\      
    \logg~(cgs) & $4.64~\pm~0.15$ \\
    \lognHE & $3.5~\pm~0.5$  \\
    \vsini~(\ensuremath{\mathrm{km~s}^{-1}}) & $90~\pm 20$  \\
    \vrad~(\ensuremath{\mathrm{km~s}^{-1}})& $-108~\pm 3$  \\
\midrule
        \multicolumn{2}{c}{SED analysis} \\
\midrule
    \ensuremath{R(55)} & $2.62~\pm~0.08$ \\
    \ensuremath{E(44-55)\,(\mathrm{mag})} & $0.109~\pm~0.02$ \\
    \ensuremath{\log(\theta(\mathrm{rad}})) & $-11.070~\pm~0.009$\\
\midrule
        \multicolumn{2}{c}{Assuming $M = 0.9\pm 0.1~M_\odot$} \\
\midrule
    \ensuremath{d_\mathrm{spec}} (kpc) & $3.9^{+0.8}_{-0.7}$ \\[3pt]
    \ensuremath{R/R_\odot} & $0.73^{+0.15}_{-0.13}$ \\[3pt]
    \ensuremath{\log L/L_\odot} & $3.57~\pm 0.18$ \\
\midrule
        \multicolumn{2}{c}{Kinematic analysis} \\
\midrule
    \ensuremath{U}~(km~s\ensuremath{^{-1}}) & $165^{+23}_{-20}$\\[3pt]
    \ensuremath{V}~(km~s\ensuremath{^{-1}}) & $7^{+56}_{-36}$\\[3pt]
    \ensuremath{W}~(km~s\ensuremath{^{-1}}) & $16^{+14}_{-16}$\\ [3pt]
    \ensuremath{e} & $0.94^{+0.05}_{-0.09}$ \\ [3pt]
    \ensuremath{L_\mathrm{z}}~(Mpc~km~s$^{-1}$) & $0.01^{+0.22}_{-0.29}$\\
\bottomrule                                            
\end{tabular}
\end{table}

\section{Radial velocity and light curve}
The SED analysis presented above rules out the presence of a luminous companion. However, a faint companion that does not contribute significantly to the SED cannot be excluded. To investigate whether HE~1518 hosts a close companion, we therefore examined the 120~s cadence light curve obtained by the Transiting Exoplanet Survey Satellite (TESS) in sectors 51 and 91, and analysed low-resolution spectroscopic observations. The additional low-resolution spectra were taken with instruments that have resolutions $\Delta\lambda$ between $3.1$ to $0.46~\AA$. The RV measurements are provided in Table~\ref{observations}. The majority of the observations were acquired in short groups spanning a few hours to several days. 

The extracted TESS light curves have CROWDSAP values of 0.85 and 0.75, indicating mild crowding but with most of the flux originating from HE~1518. We combined both sectors, and computed a Lomb–Scargle periodogram shown in Fig.~\ref{Periodogram} with a false alarm probability threshold of 0.1\% (S/N $\approx 5$) to assess the significance of any detected peaks \citep{Baran2021}. Neither in the TESS light curve nor in our RV measurements we see indications for the existence of a close binary companion. Therefore, we adopt a constant radial velocity of $108\pm3~\mathrm{km~s}^{-1}$ which is the average from the high resolution UVES and STIS spectra.

\section{Kinematic analysis}
\label{kin_sect}
We performed a kinematic analysis of HE~1518 to determine its Galactic orbit. For the Galactic potential, we adopt Model \textsc{i} from \cite{Irrgang2013}, which is a revised version of the \cite{Allen1991} potential. The estimate of our radial velocity and spectroscopic distance (see Sect.\ \ref{sect:sed}) together with right ascension, declination, and proper motion from Gaia DR3 \citep{GaiaDR3} were used to reconstruct the Galactic orbit by integrating it for 15~Gyrs. The star is on a highly eccentric orbit with an $L_\mathrm{z}$ component close to zero.

\begin{figure}
  \resizebox{\hsize}{!}{\includegraphics{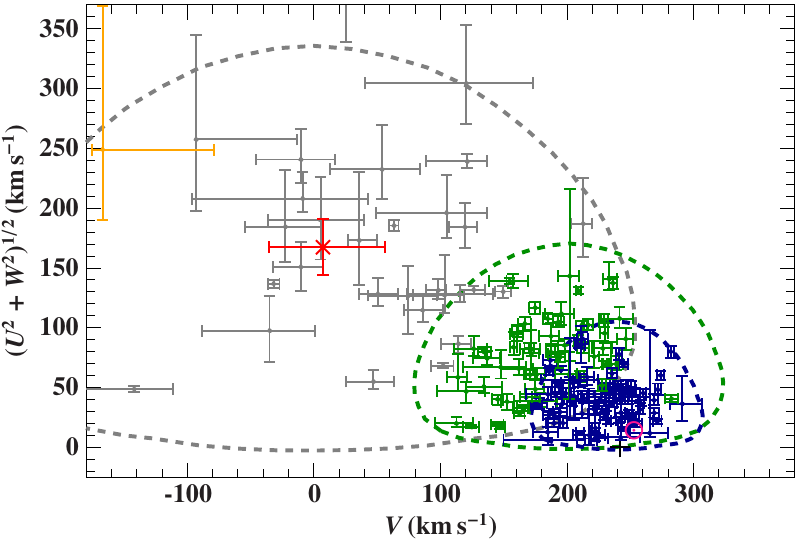}}
  \caption{Toomre diagram with HE~1518-0948 marked in red using the spectroscopic distance. In gold, we also show its position when using the parallax distance. The blue, green and grey-dotted areas represent the thin disk, thick disk and halo contours. As comparison we show the positions of the known He-sdOs from \cite{Dorsch2024}, based on the \citet{Culpan2022} sample. Their colour coding is the same as for the contours and represent their Galactic population membership. The position of the Sun is shown as a pink circle.}
  \label{Toomre}
\end{figure}

Fig.~\ref{Toomre} shows the position of the star in the Toomre diagram in red, together with the contours of the thin (blue) and thick (green) disk and the halo (grey). We also include the location that we obtained by using the parallax distance shown in orange. In both cases, the kinematic analysis indicates that HE~1518 follows a halo orbit. 

\section{Discussion}
\label{discussion}
\subsection{Mass estimate and previous evolution}\label{massdet}

\begin{figure*}
  \includegraphics[width=0.49\linewidth]{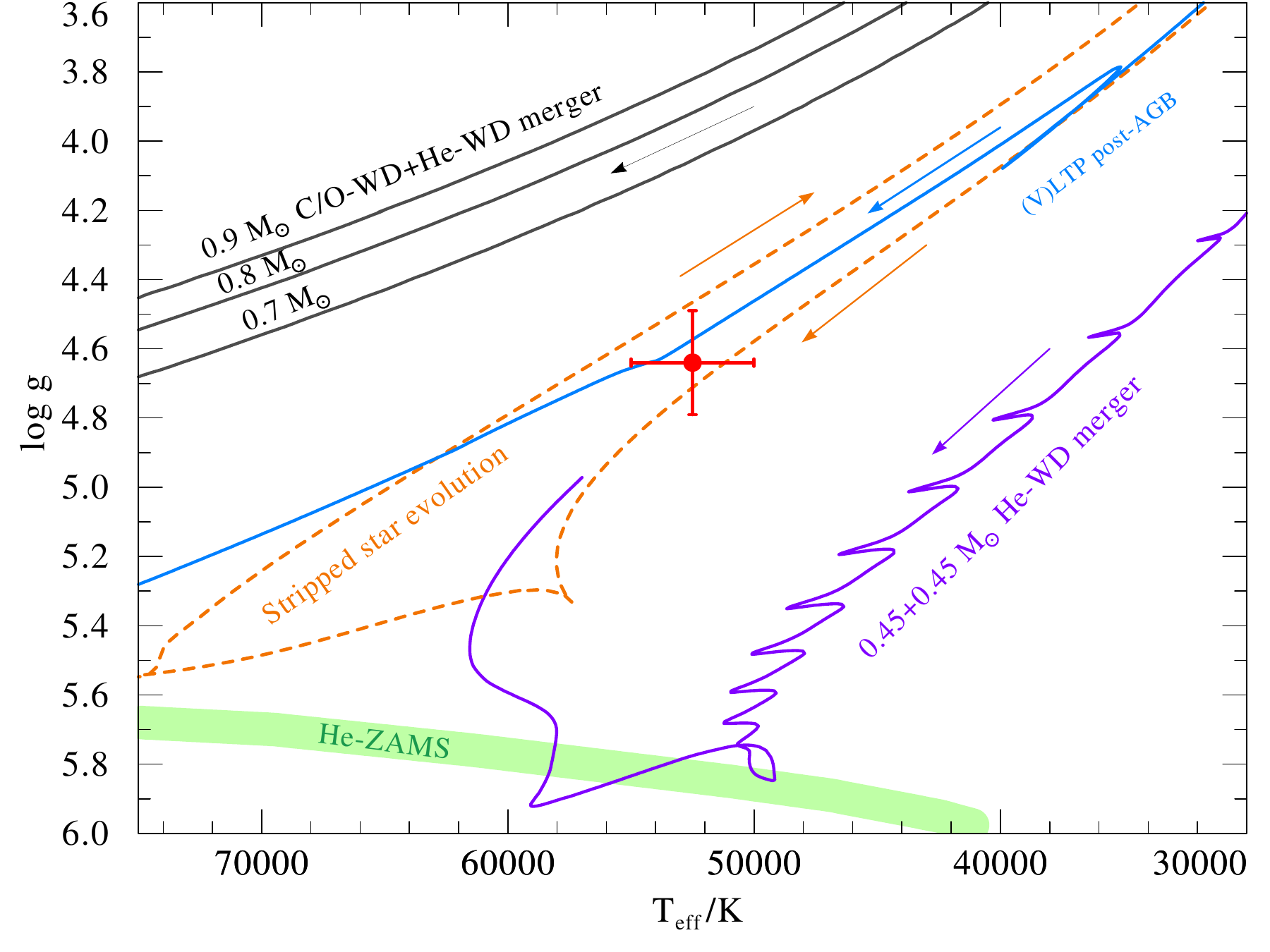}
  \includegraphics[width=0.49\linewidth]{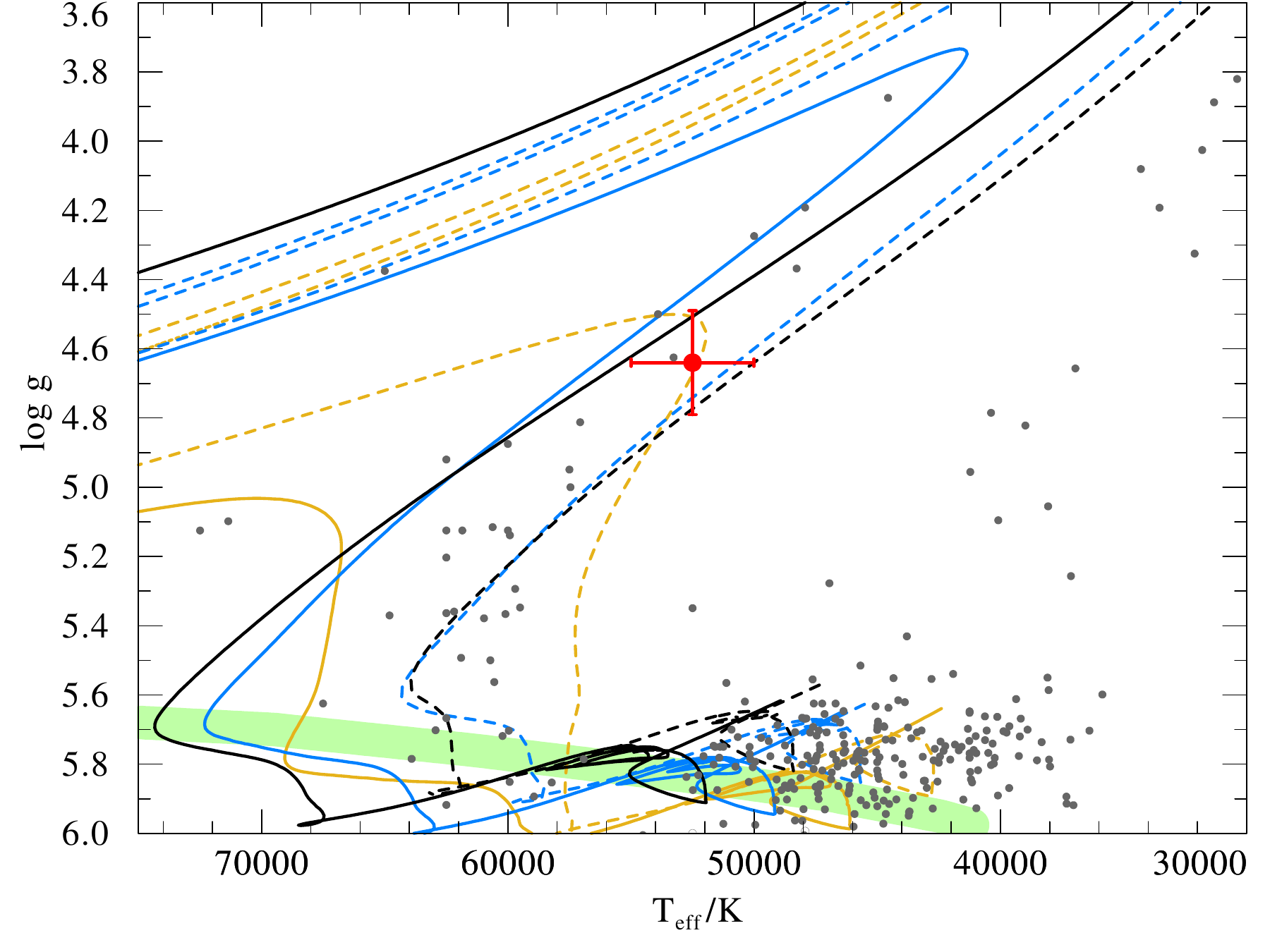}
  \caption{\textit{Left}: Kiel diagram of HE~1518 illustrating several evolutionary tracks that intersect this region of the diagram. The black solid lines show post-merger evolutionary tracks of C/O$+$He white dwarf mergers with total masses of 0.7, 0.8, and 0.9~$M_\odot$ from \citet{Saio2002}. The orange dashed line represents the evolution of an initially $7.4~M_\odot$ star stripped by a companion from \citet{Goetberg2018}, resulting in a helium star with a mass of approximately $1.9~M_\odot$. The blue line shows the evolution of a star that experienced a very late thermal pulse (VLTP) and is contracting toward higher effective temperatures to enter the white dwarf cooling track from \citet{Bertolami2006}. The model shown has a mass of $0.51~M_\odot$. The violet line corresponds to the evolution of a double helium white dwarf merger from \citet{Yu2021}, in which both components have masses of $0.45~M_\odot$. The light-green region indicates the location of the He-ZAMS from \citet{Paczynski1971} for helium stars with masses between 0.6 and approximately $2.0~M_\odot$ The arrows indicate the direction of evolution for the corresponding tracks. 
  \textit{Right}: Stellar evolution for helium stars with 0.7 (gold), 0.8 (blue) and $0.9~M_\odot$ (black) for two different metallicities $Z=0.001$ (solid line) and $Z=0.02$ (dashed line). For more details see Sect.~\ref{massdet}. The grey dots are the observed helium-rich hot subdwarfs in the SALT survey of hydrogen-deficient stars \citep[][ preliminary results]{Jeffery2026}.} 
  \label{Kiel}
\end{figure*}

Due to the uncertain parallax measurement of HE~1518, an accurate mass cannot be determined from the SED fit. Therefore, we compare the position of the star in the {\Teff}-{\logg} diagram to different theoretical tracks that cross the region where the star is located. In Fig.~\ref{Kiel} (left), we show evolutionary tracks for: a double He-WD merger with \ensuremath{0.45~M_\odot} components \citep{Yu2021}; a \ensuremath{1.9~M_\odot} stripped He-star from an initially \ensuremath{7~M_\odot} progenitor \citep{Goetberg2018}; a \ensuremath{0.8~M_\odot} post-asymptotic giant branch (pAGB) model that experienced a very late thermal pulse \citep[VLTP;][]{Bertolami2006}; and C/O-WD~+~He-WD mergers with post-merger masses of 0.7 to \ensuremath{0.9~M_\odot} \citep{Saio2002}. From these tracks, we can immediately rule out the C/O-WD~+~He-WD merger because the surface gravity of these tracks is much lower than that of our observations. 

There are scenarios in which stripped stars from intermediate to massive star progenitors can be ejected out of the Galactic thin disk \citep{Renzo2019}. In these models, the more massive primary becomes a helium star by transferring mass onto its companion. Due to the accreted mass, the secondary star evolves faster and undergoes a core-collapse supernova during the lifetime of the hot subdwarf. These models predict ejection velocities of up to $120~\mathrm{km~s}^{-1}$ for helium stars. However, this is much less than the disk-crossing velocity of   $555^{+122}_{-245}~\mathrm{km~s}^{-1}$ that we measured for HE~1518. Furthermore, the short-lived progenitors are young stars that are unlikely to have the low metallicity we obtained for HE~1518. The stripping would also expose material processed by CNO-cycle fusion, meaning that strong nitrogen rather than carbon lines should be visible. Thus, we reject this scenario. 

The only tracks left are the ones for a VLTP and a double helium WD merger. The VLTP evolutionary track from \citet{Bertolami2016} has a helium star mass of $0.5~M_\odot$. In this scenario, a delayed shell flash occurs when the star has already entered the white dwarf stage. During this event, a large fraction of the remaining helium shell is burned into carbon, producing so-called PG1159 stars. With carbon mass fractions of more than 30\% \citep{Jahn2007, Werner2016}, these stars are significantly more carbon-rich than HE~1518, for which we measure a carbon mass fraction of about 3\%. Other post-AGB scenarios create objects that still have large fractions of hydrogen, so we do not consider any of them \citep{Reindl2016, Reindl2017, Loebling2019}.  

An alternative formation channel was proposed by \citet{Justham2011}, in which an sdB star merges with a helium white dwarf companion, before becoming fully degenerate to avoid helium flashes. However, all known compact sdB+WD binaries that could merge during the sdB's evolution are located inside the Galactic disk 
\citep[e.g.][]{Kupfer2017, Kupfer2020a, Kupfer2022} and are likely originating from young, intermediate-mass progenitors. This is inconsistent with HE~1518 being an old halo star. 

Thus, only the double helium white dwarf merger scenario remains viable. For this, we must assume that both helium white dwarfs were close to their maximum possible masses prior to merging as can be seen in Fig.~\ref{Kiel} (left), in which we include a $0.45+0.45~M_\odot$ He-WD merger track from \citet{Yu2021}. However, this track terminates before reaching the observed position of the star.    

To study the He-WD merger channel in more detail, we constructed post-merger models starting as He-stars on the He-ZAMS using \textsc{LPCODE} \citep[see][and references therein]{Bertolami2022}. Convective boundary mixing (overshooting) during the core helium burning phase is defined as in \citet{Bertolami2016}. The models include a carbon mass fraction of 3\%, which is expected from the initial helium-burning flashes that lift electron degeneracy during the merger process. Two sets of tracks with masses of 0.5, 0.6, 0.7, 0.8, and 0.9~$M_\odot$ were computed for two different initial (ZAMS) metallicities ($Z=0.001$ and $Z=0.02$). These models represent the expected post-merger configuration of He-WD+He-WD mergers after the rapid phase of helium shell (sub)flashes has already ended.

The models show that, at low metallicities, only stars more massive than about $0.8~M_\odot$ become giants and go through the location of HE~1518 in the \Teff-{\logg} diagram in Fig.~\ref{Kiel} (right). Post-merger structures with lower masses evolve directly towards the white dwarf cooling sequence. He-WDs have a maximum mass of about $0.5~M_\odot$. Consequently, a He-WD+He-WD merger has a maximum theoretical mass of about $1~M_\odot$. Based on these theoretical arguments, we deduce that the mass of HE~1518 has to be $0.9\pm0.1~M_\odot$.

This establishes HE~1518 as the first hot subdwarf identified on the ascending helium giant branch and provides compelling evidence that helium-rich sdO stars form through the merger of two helium white dwarfs.
The enhanced carbon abundance in the star indicates that it was a violent merger with temperatures high enough to ignite some of the helium during this process \citep{Zhang2012, Yu2021}.  

\subsection{The future evolution of HE~1518} \label{HRDdiscussion}

The majority of hot subdwarfs have masses close to 0.5~$M_\odot$. After their core helium burning phase, they directly evolve towards hotter temperatures before entering the white dwarf cooling track. Helium stars with more than 0.8~$M_\odot$ have envelopes thick enough such that the helium burning shell around the degenerate C/O core leads to another phase of expansion where they ascend towards the helium giants. 

If HE~1518 is currently ascending the helium giant branch, it might not reach the giant phase because of its low metallicity. The track for a $0.8~M_\odot$ star is already starting to contract again after cooling to around 40~000~K. However, if it is more massive than $0.8~M_\odot$, its temperature will further decrease while the luminosity increases. When reaching a temperature range between 35~000 to 8~000~K its spectrum will resemble that of an EHe star \citep{Jeffery2008}. 

When cooling down even further, it enters the domain of hydrogen-deficient carbon stars (HdC). These stars are helium giants, which are cool enough such that carbon dust grains can condense in their outer atmosphere, releasing a highly variable dust-driven wind with high mass-loss episodes \citep{Clayton2013} as seen in R~CrB stars. Such stars are usually believed to originate from a merger of a C/O-WD+He-WD \citep{Webbink1984}, but massive double He-WD merger products like HE~1518 might also contribute to this population. 

After the end of helium shell burning, HE~1518 will rapidly evolve towards the white dwarf cooling track. During this evolution the star contracts causing the temperature to increase whereas luminosity remains nearly constant. It will evolve through a second, but now more luminous, EHe phase before being visible as an O(He) star. These stars represent the very hot, low gravity pre-WD phase before entering the WD cooling track \citep{Reindl2014}. 

\subsection{Enrichment with nuclear-processed matter}

A small number of hot subdwarfs exhibit strong enrichments of heavy elements produced by the $i$-process \citep{Dorsch2022, Battich2025}. In particular, LS~IV$-14^\circ116$ and Feige~46 (Dorsch et al. in prep.) are single stars and therefore plausible merger products. Moreover, the analysis of UV/FUSE spectra of three CN-type extreme He-sdO stars \citep{Schindewolf2018} reveals phosphorus abundances that are qualitatively consistent with the predictions of intermediate neutron captures \citep{Battich2025}. In this scenario, hydrogen is ingested into helium burning layers leading to the creation of $^{13}\mathrm{C}$ which releases a neutron when capturing an $\alpha$ particle. Notably, the only stable phosphorus isotope ($^{31}\mathrm{P}$) can only be formed by neutron captures from seed silicon nuclei or during the post core helium burning stages in massive stars. The enhanced phosphorus abundance observed in the CN-type He-sdOs studied by \citet{Schindewolf2018} is a hint that neutron capture might also occur in the formation of extreme helium sdOs such as HE~1518. Moreover, their analysis also includes one N-type He-sdO, in which this feature is not observed, as such stars are not expected to experience an episode of strong helium burning.

Motivated by these facts, we also examined the UV spectrum of HE~1518 for evidence of similar enrichment. However, at the high effective temperature of HE~1518, the relevant absorption lines are intrinsically much weaker than in cooler (\Teff$\sim35~000~\mathrm{K}$ to $40~000~\mathrm{K}$) heavy-metal subdwarfs, and the star’s rapid rotation broadens and dilutes them further. Therefore, we were unable to find clear evidence for an enrichment in heavy elements, but the upper limits provided in Table~\ref{abundances} would still allow significant enrichment.  

\begin{figure*}
  \includegraphics[width=0.49\linewidth]{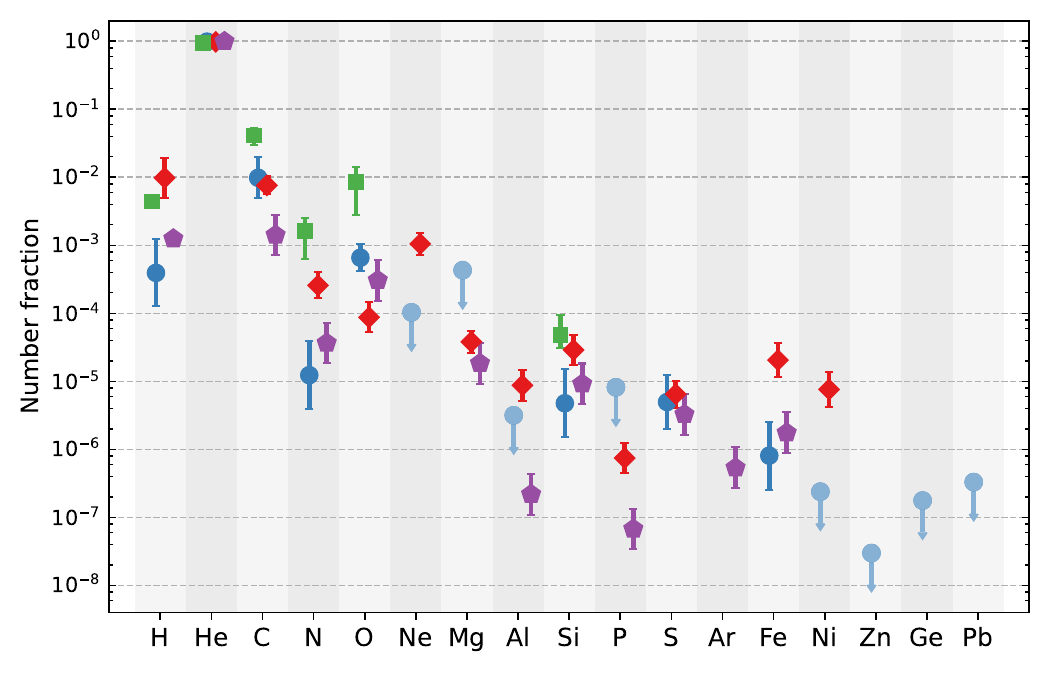}
  \includegraphics[width=0.49\linewidth]{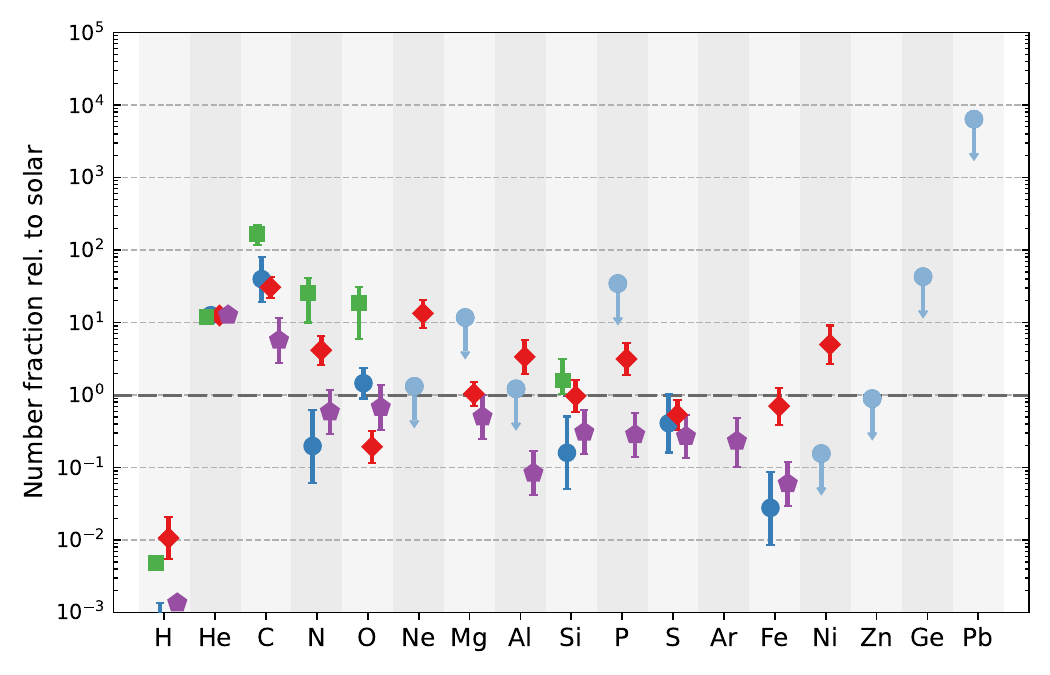}
  \caption{\textit{Left}: This plot shows the number fraction of selected elements of HE~1518 (blue points)  compared to the He-sdO $[\mathrm{CW}83]~0904-02$ from \citet{Schindewolf2018} (red diamonds), the C/O-sdO $\mathrm{UCAC}4~108-030787$ (green squares) from \citet{Werner2025} and the extreme helium star LSS~99 (violet pentagons) from \citet{Jeffery1998}. Downward pointing arrows indicate upper limits we estimated for HE~1518. 
  \textit{Right}: Same as the left plot but number fractions are shown relative to solar values.} 
  \label{comp_abu}
\end{figure*}

Compared to the CN-type objects analysed by \citet{Schindewolf2018}, HE~1518 exhibits higher carbon and oxygen abundances. In Fig.~\ref{comp_abu} we compare our abundance measurements of HE~1518 to the CN-type He-sdO $[\mathrm{CW}83]~0904-02$ from \citet{Schindewolf2018}. Models of equal-mass mergers by \citet{Zhang2012} predict increasing carbon abundances with higher total masses and lower metallicities. In these models, most of the nitrogen is converted into $^{18}\mathrm{O}$ and $^{22}\mathrm{Ne}$ via $\alpha$-capture reactions. Thus, a significant fraction of the nitrogen in HE~1518 either survived the merger or was subsequently re-accreted from a disk. The detection of nitrogen in HE~1518 therefore raises the possibility that purely carbon-rich (C-type) He-sdOs may not exist. In this case, the absence of nitrogen in the optical spectra of some He-sdOs may reflect an intrinsically low metallicity rather than the outcome of an uncontaminated hot merger. 

We were unable to detect neon in HE~1518, although our upper limit lies well below the abundances measured by \citet{Schindewolf2018}. Their analysis also revealed a clear correlation between nitrogen and neon, with ratios $1.6<n_\mathrm{Ne}/n_\mathrm{N}<4$. The expected neon abundance in HE~1518 may therefore fall below the detection threshold. 

In Fig.~\ref{comp_abu} we also compare HE~1518 to $\mathrm{UCAC}4~108-030787$ (hereafter $\mathrm{UCAC}4~108$), which belongs to the class of recently discovered C/O-sdOs \citep{Werner2025}. These objects are created when a less massive He/C/O hybrid white dwarf merges with a more massive helium white dwarf. The carbon- and oxygen-rich material is then visible on the surface of the merger product \citep{Bertolami2022}. Only three of these objects have been observed so far \citep{Werner2022, Werner2025}. In comparison to HE~1518 the stars have similar temperatures but higher surface gravities and lower luminosities. Of these stars $\mathrm{UCAC}4~108$ has the lowest carbon and oxygen fraction, but they are still three times larger compared to HE~1518.

The star LSS~99 belongs to the luminous class of extreme helium stars \citep{Monai2024}, which are thought to originate from C/O-WD+He-WD mergers. It was analysed by \citet{Jeffery1998} and is also a metal-poor object. Its abundance pattern is shown in Fig.~\ref{comp_abu}. Among the comparison objects, LSS~99 shows the closest agreement with the abundance pattern of HE~1518, although its carbon and oxygen abundances are lower. As in HE~1518, silicon and sulphur are enhanced relative to iron, which is typical in metal-poor stars. Unlike the CN-type He-sdO $[\mathrm{CW}83]~0904-02$, LSS~99 does not exhibit a phosphorus overabundance relative to silicon and sulphur. 

In contrast to the known heavy-metal hot subdwarfs, stars such as HE~1518 may have the potential to release the created elements into the interstellar medium (ISM). If HE~1518 reaches the R~CrB stage, the heavy elements would be expelled by the dust-driven wind that can reach mass-loss rates of up to $10^{-6}\,M_\odot~\mathrm{yr}^{-1}$ \citep{Clayton2013}. However, there also exists a population of dustless HdC (dLHdC) stars that have the same spectral characteristics as R~CrB stars but appear less luminous \citep{Tisserand2022} and, due to their lack of dust, show no evidence for strong mass loss \citep{Geballe2009}. Whether these objects represent a permanent evolutionary state or a quiescent phase during R~CrB evolution is currently unknown, although some exhibit indications of recent dust formation \citep{Tisserand2022}.

In our $0.9~M_\odot$ model, the R~CrB phase lasts approximately $100~\mathrm{kyr}$, comparable to the lifetimes of R~CrB models produced by C/O-WD+He-WD mergers in \citet{Crawford2024}. Over this period, such stars can return up to $0.1~M_\odot$ of material, that is potentially enriched, to the ISM. Some R~CrB and dLHdC stars show strong enrichments of Sr, Y, and Ba that can only be created during the merger process because, afterwards, the convective envelope is disconnected from the helium buring shell \citep{Crawford2022}. If the $i$-process can also occur in C/O-WD+He-WD mergers, then R~CrB stars could be a significant source in the production of heavy elements. A detailed study is beyond the scope of this paper, but is a compelling topic for future research. 

\subsection{Why is HE~1518 rotating so fast?}
One of the distinctive features of HE~1518 is the fast projected rotational velocity of the star of $90\pm20~\mathrm{km~s^{-1}}$. This is even more remarkable because the star expanded already and therefore has a radius that is around three times larger compared to its core helium burning radius. Thus, assuming angular momentum conservation, the star would have rotated at $\sim200$–$300~\mathrm{km~s^{-1}}$ on the He-ZAMS, substantially faster than the majority of He-sdOs analysed by \citet{Hirsch2009}. For their sample, typical projected rotational velocities of ${v\sin i}=20$–$30~\mathrm{km~s^{-1}}$ were measured, in agreement with theoretical He-WD+He-WD merger models by \citet{Schwab2018}. In their sample of 34 He-sdOs they report a projected rotational velocity above 100~km~s$^{-1}$ only for one star (PG~1632+223). Therefore, only a small fraction of He-sdOs seem to be very fast rotators. 
This also includes the ejected hot subdwarf US\,708 which was likely spun up in a very close binary and not formed by a merger \citep{Geier2015}. 

C-type objects are expected to form predominantly through hot mergers. In this picture, the merger of two helium white dwarfs of nearly equal mass should provide the most favourable conditions for producing a carbon-rich remnant. This might also provide a potential explanation for the high rotational velocity of HE~1518. Theoretical predictions suggest that in nearly equal mass mergers the resulting accretion disk around the newly formed object might only contain small amounts of mass \citep{Aguilar2009, Dan2014}, which would help to preserve a high rotational velocity because angular momentum is transported outward in the disk, causing the stars to spin down. 

This is in agreement with the results of \citet{Hirsch2009} who found that the N-type He-sdOs have on average a smaller rotational velocity compared to the the CN- and C-type objects. The N-type He-sdOs are created in slow merger events \citep{Zhang2012} where the material is accreted from a disk created by the disruption of the lower mass star. \citet{Hirsch2009} were able to measure a rotational velocity of more than $10~\mathrm{km~s}^{-1}$ only in one third of the N-type objects, which was their detection limit.

\subsection{Potential misinterpretations}

Due to its high effective temperature and luminosity, the position of HE~1518 in the HRD overlaps with that expected for more massive hot subdwarfs formed through the stripping of intermediate-mass stars. Indeed, objects with spectra similar to HE~1518 have recently been identified in the Large Magellanic Cloud by \citet{Drout2023} and \citet{Goetberg2023}, and are most likely also ascending the helium giant branch. Consequently, massive He-WD+He-WD and C/O-WD$+$He-WD mergers should be considered when interpreting populations of apparently more massive hot subdwarfs created through binary stripping. These merger products can be distinguished by their extreme helium enrichment, absence of detectable hydrogen, and strong carbon features, whereas binary-stripped stars such as the massive He-sdO+WD binary HD~49798 typically retain significant hydrogen and show a CNO burning pattern (N-rich, C/O-poor) \citep{Rauch2025}.  

Furthermore, the star might also be confused as a runaway. With a disk-crossing velocity of $555^{+122}_{-245}~\mathrm{km~s}^{-1}$ and its high rotational velocity, it resembles the predicted properties of type~Ia supernova–ejected hot subdwarfs described by \citet{Neunteufel2022}. Only its low metallicity reveals that the star belongs to the intrinsic halo population.

These examples show that stars like HE~1518 might be misidentified as different objects if abundances are not taken into account.

\section{Summary and Conclusions}
We performed a detailed analysis of the luminous He-sdO HE~1518. A comparison of its position in the \teff–\logg~diagram with evolutionary tracks for stripped binary stars, VLTP objects, C/O-WD+He-WD, and He-WD+He-WD merger scenarios indicates that a double He-WD merger provides the most plausible formation channel, despite the star’s unusually low \logg~relative to most known He-sdOs. The evolutionary tracks reveal that HE~1518 is currently undergoing helium shell burning causing it to ascend the helium giant branch. With an inferred mass of at least $0.8\,M_\odot$, it is close to the upper mass limit that a double He-WD merger can reach. In comparison to the majority of known He-sdOs, the star has an outstandingly high rotational velocity.

We derived accurate abundances for He, C, N, O, Si, S, and Fe, and determined upper limits for H, Ne, Mg, P, Ar, Ti, Cr, Mn, Co, Ni, Zn, Ge, Zr, Sr, and Pb by fitting optical and UV spectra. The resulting chemical composition indicates a very low metallicity, consistent with the halo orbit of HE~1518. Due to its high effective temperature and rapid rotation, we were unable to identify signatures of $i$-process nucleosynthesis that is observed in a few intermediate helium-rich hot subdwarfs. The star is enriched in carbon and oxygen similar to other CN-type He-sdOs. It shows signatures of a weak stellar wind making it a unique helium-rich object in an extremely low metallicity environment well suited to study wind models. 

To further constrain the formation channel of HE~1518-like stars, a  larger sample is required, enabling a systematic comparison with the population of double He-WD binaries. This  progenitor population also has to be constrained. We identify two candidate double He-WD systems, WD~0455$-$295 \citep{Napiwotzki2020} and Gaia~DR3~928585329693880832 \citep{Brown2022}, that are expected to merge within a Hubble time and may thus form a He-sdO  massive enough to  ascend the helium giant branch. Similar WD binaries have been reported by \citet{Munday2025}, although their longer orbital periods imply merger timescales exceeding a Hubble time.

\begin{acknowledgements}
MD was supported by the Deutsches Zentrum für Luft- und Raumfahrt (DLR) through grant 50-OR-2510. 
MP received funding by the Deutsche Forschungs-
gemeinschaft (DFG) through grants GE2506/9-1, GE2506/12-1, and GE2506/18-1. HD was supported by the DFG through grant  GE2506/9-2.

Part of this work was supported by the
CONICET-DAAD 2022 bilateral cooperation grant number 80726. M3B is partially funded by CONICET and Agencia I+D+i through grants PIP-2971 and
PICT 2020-03316. 

Based on observations made with the \textit{Isaac Newton} and \textit{William Herschel} telescopes operated on the island of La Palma by the Isaac Newton Group of Telescopes in the Spanish Observatorio del Roque de los Muchachos of the Instituto de Astrofísica de Canarias.

Based on observations made with ESO Telescopes at the La Silla Paranal Observatory under programme IDs 167.D-0407(A) and 71.D-0380(A).

Based on observations obtained at the Southern Astrophysical Research (SOAR) telescope, which is a joint project of the Minist\'{e}rio da Ci\^{e}ncia, Tecnologia e Inova\c{c}\~{o}es (MCTI/LNA) do Brasil, the US National Science Foundation’s NOIRLab, the University of North Carolina at Chapel Hill (UNC), and Michigan State University (MSU).

This research is based on observations made with the NASA/ESA Hubble Space Telescope obtained from the Space Telescope Science Institute, which is operated by the Association of Universities for Research in Astronomy, Inc., under NASA contract NAS 5–26555. These observations are associated with program A far-UV treasury survey for hot subdwarf stars (ID 17697; PI: M.Dorsch)
\end{acknowledgements}

\bibliographystyle{aa}
\bibliography{sample}

\begin{appendix}

\section{Additional tables and figures}

\begin{table*}
\caption{Observations of HE~1518-0948}
\label{observations}      
\centering                                      
\begin{tabular}{c c c c c c}          
\toprule\toprule                      
Observation date & MJD & Telescope & Instrument & $\Delta\lambda[\AA]$ & Radial velocity $[\mathrm{km~s}^{-1}]$ \\ \midrule
23/04/2002  & 52387.3560 & ESO-VLT & UVES& $R=20000$ & $-107~\pm3$\\ 
12/05/2003  & 52771.2680 & ESO-VLT & UVES& $R=20000$ & $-103~\pm3$\\  
14/02/2010  & 55241.3510 & SOAR & Goodman &2.0 & $-99\pm10$\\ 
14/02/2010  & 55241.3550 & SOAR & Goodman &2.0 & $-98\pm10$\\
14/02/2010  & 55241.3590 & SOAR & Goodman  &2.0 & $-98\pm10$\\
14/02/2010  & 55241.3650 & SOAR & Goodman  &2.0 & $-101\pm10$\\
14/02/2010  & 55241.3690 & SOAR & Goodman &2.0 & $-101\pm10$\\
14/02/2010  & 55241.3730 & SOAR & Goodman &2.0 & $-100\pm10$\\
14/02/2010  & 55241.3790 & SOAR & Goodman &2.0 & $-101\pm10$\\
14/02/2010  & 55241.3830 & SOAR & Goodman &2.0 & $-96\pm10$\\
26/06/2010  & 55373.9664 & INT & IDS& 3.1 & $-95\pm15$\\
26/06/2010  & 55373.9748 & INT & IDS& 3.1 & $-107\pm15$\\
26/06/2010  & 55373.9831 & INT & IDS& 3.1 & $-103\pm15$\\
26/06/2010  & 55373.9914 & INT & IDS& 3.1 & $-125\pm15$\\
26/06/2010  & 55374.0010 & INT & IDS& 3.1 & $-111\pm15$\\
26/06/2010  & 55374.0101 & INT & IDS& 3.1 & $-112\pm15$\\
26/06/2010  & 55374.0211 & INT & IDS& 3.1 & $-112\pm15$\\
26/06/2010  & 55374.0296 & INT & IDS& 3.1 & $-124\pm15$\\
29/06/2010  & 55376.9318 & INT & IDS& 3.1 & $-128\pm15$\\
29/06/2010  & 55376.9429 & INT & IDS& 3.1 & $-115\pm15$\\
29/06/2010  & 55376.9524 & INT & IDS& 3.1 & $-99\pm15$\\
29/06/2010  & 55376.9607 & INT & IDS& 3.1 & $-111\pm15$\\
29/06/2010  & 55376.9707 & INT & IDS& 3.1 & $-110\pm15$\\
11/08/2011  & 55784.0135 & SOAR & Goodman& 0.46 & $-106\pm10$\\
11/08/2011  & 55784.0620 & SOAR & Goodman& 0.46 & $-99\pm10$\\
03/06/2013  & 56446.0303 & WHT & ISIS& 1.1 & $-88\pm 10$\\
04/06/2013  & 56447.0233 & WHT & ISIS& 1.1 & $-102\pm 10$\\
05/06/2013  & 56448.0061 & WHT & ISIS& 1.1 & $-106\pm 10$\\
06/06/2013  & 56449.0486 & WHT & ISIS& 1.1 & $-87\pm 10$\\
12/08/2013  & 56516.9211 & WHT & ISIS& 1.1 & $-91\pm 10$\\
13/08/2013  & 56517.8796 & WHT & ISIS& 1.1 & $-93\pm 10$\\
13/08/2013  & 56517.8819 & WHT & ISIS& 1.1 & $-117\pm 10$\\  
10/01/2025  & 60685.9686 & HST & STIS & $R=45800$ & $-113~\pm3$\\
\bottomrule                                            
\end{tabular}
\end{table*}

\begin{table}
\caption{Lines used to determine upper abundance limits. }
\label{elements}      
\centering                                      
\begin{tabular}{l l}          
\toprule\toprule                     
Ion & Wavelength [\AA] \\ \midrule
Ne \textsc{iii}  &  1295.504, 1460.107, 1461.237 \\
Mg \textsc{iii} &  1447.264, 1572.713, 1697.274 \\
Mg \textsc{iv}  &  1698.788 \\
Al \textsc{iii}  &  1605.766, 1611.873 \\
P \textsc{v}    & 1447.87, 1610.50 \\
Ar \textsc{v}  &  1202.401, 1350.424 \\
Ar \textsc{vi}  &  1283.918, 1303.856, 1324.409, 1325.729 \\
Ti \textsc{iv}  &  1451.739, 1467.343, 1469.191 \\
Cr \textsc{v}  &  1193.954, 1465.859, 1481.655, 1482.756 \\
Cr \textsc{v}  & 1484.662, 1489.705, 1497.972 \\
Cr \textsc{v}  & 1519.032, 1579.694, 1603.199 \\
Mn \textsc{v}  & 1359.235, 1436.930, 1437.387  \\
Mn \textsc{v}  & 1443.308, 1446.493, 1448.132   \\
Mn \textsc{v}  & 1482.514, 1483.070, 1486.498   \\
Mn \textsc{v}  & 1503.581, 1504.343, 1516.786 \\
Co \textsc{v}  &  1275.953, 1277.022, 1281.673 \\
Co \textsc{v}  &  1331.017, 1331.499, 1413.420 \\
Ni \textsc{v}  & 1244.174, 1250.388, 1251.812 \\
Ni \textsc{v}  & 1261.760, 1273.204, 1300.979 \\
Ni \textsc{v}  & 1311.106, 1318.515, 1336.136 \\
Zn \textsc{iv}  & 1306.657, 1320.704, 1321.215, 1322.428 \\
Zn \textsc{v}  & 1174.346, 1176.868, 1185.898 \\
Ge \textsc{iv}  &  1089.491, 1229.840 \\
Sr \textsc{iv}  & 1331.13, 1347.90, 1356.22, 1361.16  \\
Sr \textsc{v}  &  1311.78, 1372.59, 1379.615 \\
Sr \textsc{v}  &  1380.48, 1396.31, 1412.958, 1415.404 \\
Zr \textsc{iv}  & 1598.95\\
Zr \textsc{v}  &  1303.93, 1323.83, 1332.07\\
Pb \textsc{v} & 1157.879,  1189.953,  1197.700\\
Pb \textsc{v}  &  1233.500, 1248.460 \\
\bottomrule                                            
\end{tabular}
\end{table}

\begin{figure}[!h]
  \resizebox{\hsize}{!}{\includegraphics{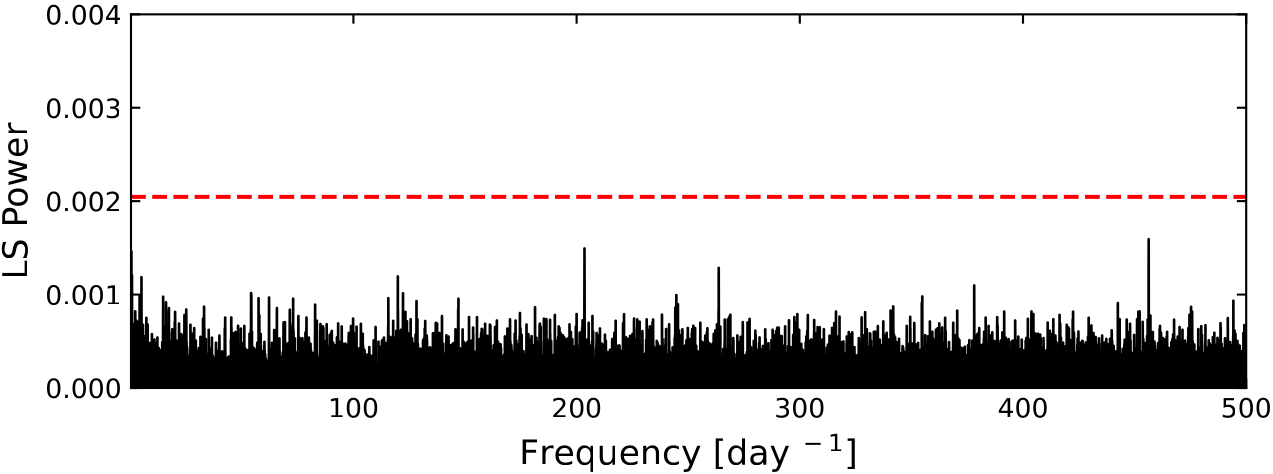}}
  \caption{Periodogram of HE1518. The red-dotted line outlines our FAP threshold corresponding to a S/N of $\approx$~5. Aliasing effects caused by gaps in the time-series data dominate the low frequency regime corresponding to periods longer than 3 days, and we therefore truncate the periodogram accordingly. These results show no indication of variability above the significance threshold.}
  \label{Periodogram}
\end{figure}

\begin{figure}[!h]
\centering
\includegraphics[width=0.99\columnwidth]{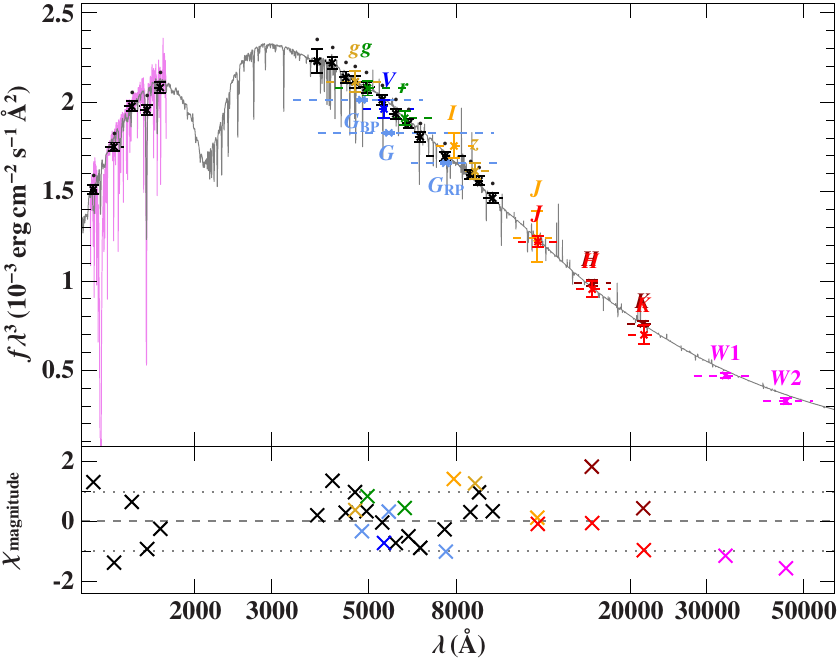}
\caption{The SED fitting was performed using photometry from the following surveys covering the UV to infrared range: the HST/STIS spectrum, \textit{Gaia} XP spectra and $G$/$G_{BP}$/$G_{RP}$ photometry \citep{GaiaBPRP2022}, VISTA \citep{VISTA1, VISTA2}, VST \citep{VST2015}, Skymapper \citep{Skymapper2024}, DENIS \citep{Denis2005}, 2MASS \citep{2Mass2003}, and WISE \citep{WISE1, WISE2, WISE3}.}
\label{SED}
\end{figure}

\begin{figure}[!h]
\centering
\includegraphics[width=0.49\textwidth]{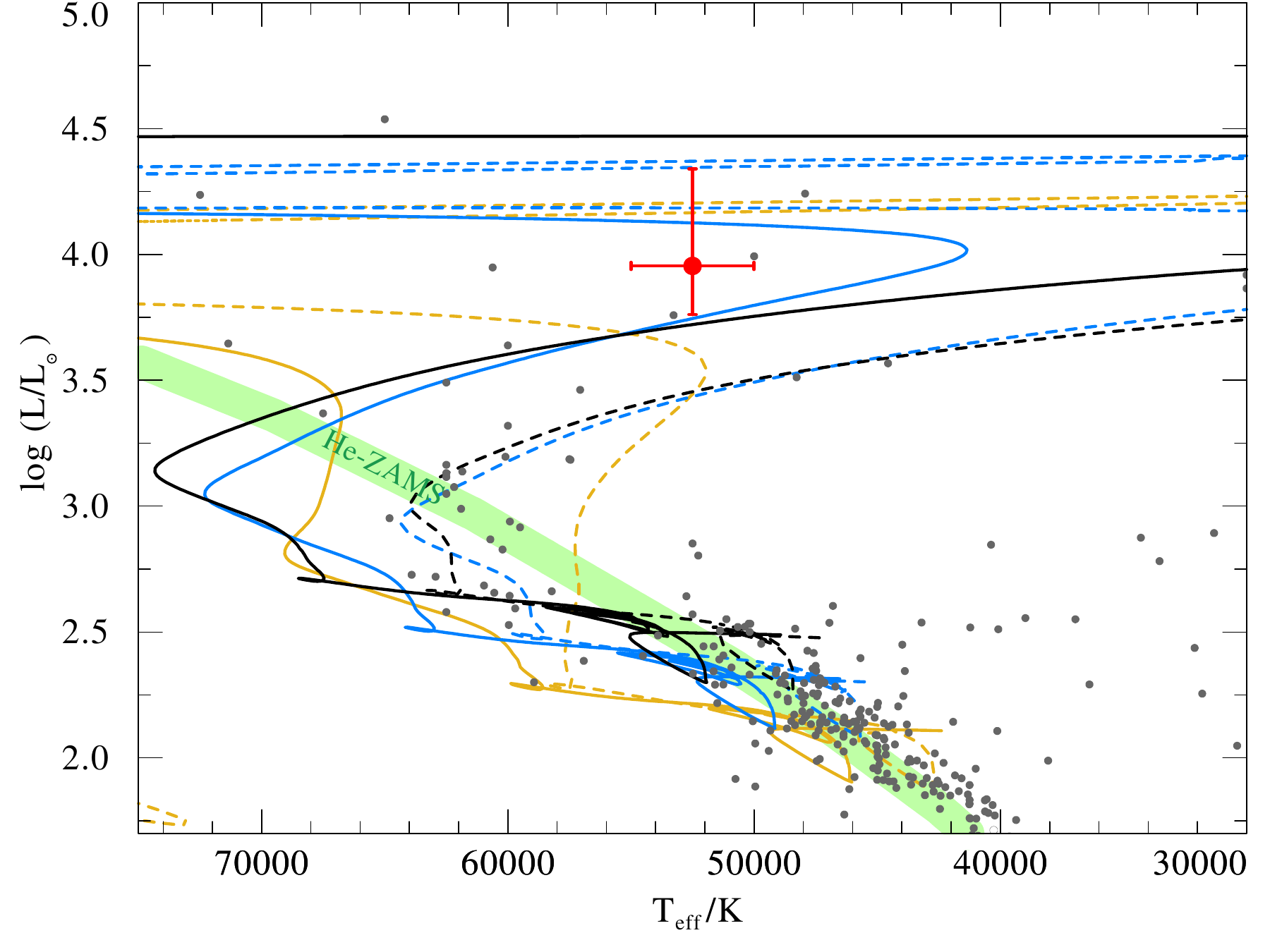}
\caption{The plot shows the position of HE~1518 in the HRD. The tracks and meaning of the colours is the same as in Fig.~\ref{Kiel} (right).}
\label{SED_fig}
\end{figure}

\end{appendix}

\end{document}